\documentclass[reprint,prl,superscriptaddress,amsmath,amssymb,aps,floatfix,nofootinbib]{revtex4-2}
\usepackage[pdftex]{graphicx}
\usepackage{bm}
\usepackage{color}
\usepackage{times}
\usepackage{amsmath}
\usepackage{amssymb}
\usepackage{natbib}
\usepackage{float}

\begin{document}

\title{Ostwald Ripening in Underground Gas Storage}

\author{Mohammad Salehpour}
\affiliation{Department of Civil Engineering, McMaster University, Hamilton, ON, Canada}
\author{Tian Lan}
\affiliation{Department of Civil Engineering, McMaster University, Hamilton, ON, Canada}
\author{Nicolas Bueno}
\affiliation{Department of Energy and Mineral Engineering, The Pennsylvania State University, University Park, PA, USA}
\author{Md Zahidul Islam Laku}
\affiliation{Department of Energy and Mineral Engineering, The Pennsylvania State University, University Park, PA, USA}
\author{Yashar Mehmani}
\affiliation{Department of Energy and Mineral Engineering, The Pennsylvania State University, University Park, PA, USA}
\author{Benzhong Zhao}
\email{robinzhao@mcmaster.ca}
\affiliation{Department of Civil Engineering, McMaster University, Hamilton, ON, Canada}

\begin{abstract}
Underground gas storage is central to both climate mitigation and energy transition strategies, supporting both long-term carbon sequestration and seasonal hydrogen storage. A key mechanism governing the fate of injected gases is Ostwald ripening---the curvature-driven mass transfer between trapped gas ganglia in porous media. While ripening is well understood in open systems, its behavior in geometrically confined porous structures remains poorly characterized, especially over long timescales relevant to subsurface operations. Here, we present ultra-high-resolution microfluidic experiments that capture the evolution of residually trapped hydrogen over weeks in realistic, heterogeneous pore geometries under well-defined boundary conditions. We observe a distinct two-stage dynamic: a rapid local equilibration among neighboring bubbles, followed by slow global depletion driven by long-range diffusion toward low–chemical potential boundaries. Building on these insights, we develop and validate a continuum model that couples microscale capillary pressure–saturation ($P_c$–$S$) relationships, extracted via the pore-morphology method, with macroscopic diffusive transport. The model accurately predicts gas saturation evolution without fitting parameters and collapses experimental results across a range of experimental conditions. Extending the model to reservoir scales, we estimate equilibration timescales ($t_{\text{eq}}$) for CO$_2$ and H$_2$ in homogeneous sandstone aquifers. We find that ripening occurs much faster than convective dissolution in CO$_2$ sequestration, and on timescales comparable to seasonal H$_2$ storage operations. These findings establish a quantitative framework linking pore-scale heterogeneity to field-scale gas redistribution, with implications for the design and longevity of subsurface storage strategies.
\end{abstract}

\maketitle

\section{Introduction}

Underground gas storage is essential to addressing contemporary energy and environmental challenges~\cite{krevor2023subsurface}. Geologic sequestration of carbon dioxide (CO$_2$) provides a long-term solution for reducing atmospheric greenhouse gas concentrations by permanently storing CO$_2$ in deep geological formations~\cite{lackner2003,szulczewski-pnas-2012}. Concurrently, seasonal underground hydrogen (H$_2$) storage offers a scalable means of buffering the intermittency of renewable energy sources, thereby facilitating the reliable integration of low-carbon energy into the broader energy infrastructure~\cite{heinemann2021}.

As gases such as CO$_2$ and H$_2$ are injected into the subsurface, they displace the wetting ambient brine in a drainage process. Upon subsequent reinvasion of the brine---referred to as imbibition---portions of the injected gas become immobilized as discrete ganglia surrounded by the wetting phase, in a process known as residual trapping. In geological carbon sequestration, residual trapping primarily occurs as buoyant CO$_2$ migrates upward~\cite{krevor-ijggc-2015}. In underground hydrogen storage, it arises from the cyclical drainage and imbibition of the pore space during repeated injection and withdrawal cycles~\cite{thaysen2023pore}.

Once trapped, these isolated gas ganglia partially dissolve into the surrounding aqueous phase. Over time, molecular diffusion drives mass exchange between bubbles of differing sizes and curvatures---a process known as Ostwald ripening~\cite{ostwald1897studien,voorhees1985theory,fu2016thermodynamic}. This redistribution of mass causes some bubbles to shrink and eventually vanish, while others to grow sufficiently to overcome capillary constraints, regain buoyancy, and migrate upward to rejoin the mobile gas plume. Such dynamics may prove advantageous for underground hydrogen storage by reducing residual hydrogen losses; however, they pose a significant concern in geological carbon sequestration, as bubble mobilization can increase the risk of CO$_2$ leakage. In addition, ripening-induced redistribution of bubbles modifies local flow and chemical conditions, thereby influencing geochemical reactions such as rock dissolution and mineral precipitation~\cite{jimenez2020homogenization}.

Ostwald ripening, first described in the late 19th century, has been extensively studied in open systems, where gas bubbles grow unconstrained by geometry. Its investigation within porous media, however, is a more recent development. Pore-scale observations have revealed that geometric confinement fundamentally controls the ripening dynamics. In micromodels with homogeneous pore structures, an initially polydisperse population of gas bubbles evolves toward a monodisperse state, with bubbles attaining uniform size~\cite{xu2017egalitarianism}. This behavior changes markedly in the presence of pore-scale heterogeneity. Pore-scale modeling studies predict that under such heterogeneity, ripening leads to a stable equilibrium distribution of bubble sizes, governed by the interplay between initial conditions and the underlying pore geometry~\cite{dechalender2018pore,mehmani2022pore,singh2024ostwald}. Recent X-ray computed tomography (CT) experiments of hydrogen storage in rock cores reveal that ripening behavior is dictated by ganglion size: medium-sized ganglia fragment, larger ganglia grow, and smaller ones dissolve completely~\cite{zhang2023pore,jangda2023pore,boon2024multiscale}. These CT observations, however, are typically limited to small subdomains and short timeframes ($\sim$1 day), providing only a partial---and potentially unrepresentative---snapshot of the ripening process. Concurrently, continuum-scale models have been developed that couple capillary pressure–saturation ($P_c$–$S$) relationships with Henry’s law to describe the macroscopic evolution of Ostwald ripening in porous media~\cite{li2020continuum,mehmani2024continuum}. However, the appropriate choice of $P_c$–$S$ relationships for modeling ripening remains unclear, and the predictive accuracy of these models has yet to be tested against experimental data. A promising approach for extracting $P_c$–$S$ relationships from microscale geometry was recently proposed by Mehmani and Xu~\cite{mehmani2024continuum}, and is validated in this work.

In this study, we perform microfluidic experiments to investigate Ostwald ripening of hydrogen in realistic pore geometry under well-defined boundary conditions. By achieving ultra-high-resolution imaging (0.3~$\mu$m/pixel) across the entire microfluidic domain, we track the collective evolution of a large bubble population while simultaneously resolving the interface curvature---and thus the capillary pressure---of individual bubbles with high accuracy. Each experiment spans over 15 days, an order of magnitude longer than in previous studies, allowing us to capture long-term dynamics. We observe that neighboring bubbles rapidly reach local equilibrium, as pore-scale heterogeneities provide a multitude of metastable configurations that facilitate equilibration through Ostwald ripening. In contrast, equilibration with the boundaries---which are much larger than a typical pore throat---occurs significantly more slowly, governed by the timescale of long-range diffusive mass transfer.

Crucially, we demonstrate for the first time that a continuum model incorporating microscale pore-geometry characteristics can accurately predict the evolution of gas saturation without the use of fitting parameters. When applied to representative sandstone formations, the model yields characteristic equilibration times for CO$_2$, H$_2$, and CH$_4$, showing that Ostwald ripening proceeds significantly faster than convective dissolution in carbon sequestration scenarios and occurs on timescales comparable to seasonal hydrogen storage cycles. These findings establish the first validated, predictive framework that links pore-scale heterogeneity to field-scale gas redistribution.

\section*{Microfluidic Experiments}

\begin{figure}
	\centering
	\includegraphics[width=8.7cm]{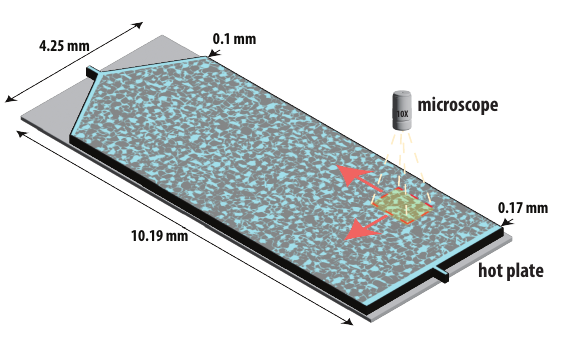}
    \caption{We conduct microfluidic experiments of hydrogen Ostwald ripening in micromodels patterned with realistic pore geometries and well-defined boundary conditions. Ultra-high-resolution imaging (0.3~$\mu$m/pixel) is achieved by scanning the entire micromodel in a serpentine pattern using overlapping sub-windows (indicated by dashed lines). \label{fig:experiment}}
\end{figure}

We conduct microfluidic experiments to investigate Ostwald ripening among residually trapped hydrogen bubbles in porous media. The micromodel is fabricated from a two-dimensional X-ray CT image of a Canadian sandstone, preserving the pore-scale heterogeneities characteristic of natural porous media in the subsurface (Fig.~\ref{fig:experiment}). The resulting structure exhibits a median pore size of 42~$\mu$m, a median pore-throat size of 10.6~$\mu$m, and a porosity of 0.3 (\emph{SI Appendix, Fig.~S1}). The micromodel is laterally bounded by inlet and outlet channels with widths of 100~$\mu$m and 170~$\mu$m, respectively. The device is fabricated by Interface Fluidics (Canada) using deep reactive ion etching (DRIE) to a depth of 7.8~$\mu$m on a silicon wafer, followed by anodic bonding to borosilicate glass. This approach provides high mechanical integrity and hermetic sealing, effectively preventing hydrogen leakage during long-term experiments. A heating element is placed beneath the micromodel to maintain uniform and controlled temperature conditions throughout the experiments.

To initiate each experiment, the micromodel is first saturated with deionized water (MilliporeSigma). Research-grade hydrogen gas (Air Liquide) is then introduced at a constant rate for a total of 50 pore volumes. Subsequently, water is reinjected for a total of 100 pore volumes, resulting in the residual trapping of hydrogen~\cite{buchgraber2012} with an initial gas saturation $S_g^\text{init}$ ranging from 0.49 to 0.65. To establish well-defined boundary conditions, hydrogen bubbles are deliberately retained in both the inlet and outlet channels, which are significantly larger than the pore throats in the porous matrix. These larger bubbles exhibit lower chemical potential, thereby driving diffusive mass transfer from the porous matrix toward the boundaries. The system is allowed to equilibrate until high-precision single-crystal quartz resonant pressure sensors (XTALX LUQS1, Phase Sensors) at both the inlet and outlet return to atmospheric pressure. Once pressure equilibrium is reached, the inlet and outlet valves are closed, thereby isolating the micromodel and initiating the Ostwald ripening process. Experiments are conducted at two distinct temperatures, 40$^\circ$C and 80$^\circ$C, with each experiment lasting 24 days and 15 days, respectively, except in one case where the experiment was terminated prematurely due to an overnight power outage. The micromodel is mounted on the motorized stage of an upright microscope (BX51, Olympus), and time-lapse imaging is performed by scanning the domain in a serpentine pattern. This imaging protocol yields an ultra-high resolution of 0.3~$\mu$m/pixel, with each full-frame image composed of 120 stitched sub-images~(\emph{Materials and Methods}).

\begin{figure*} [t!]
	\centering
	\includegraphics[width=5.5in]{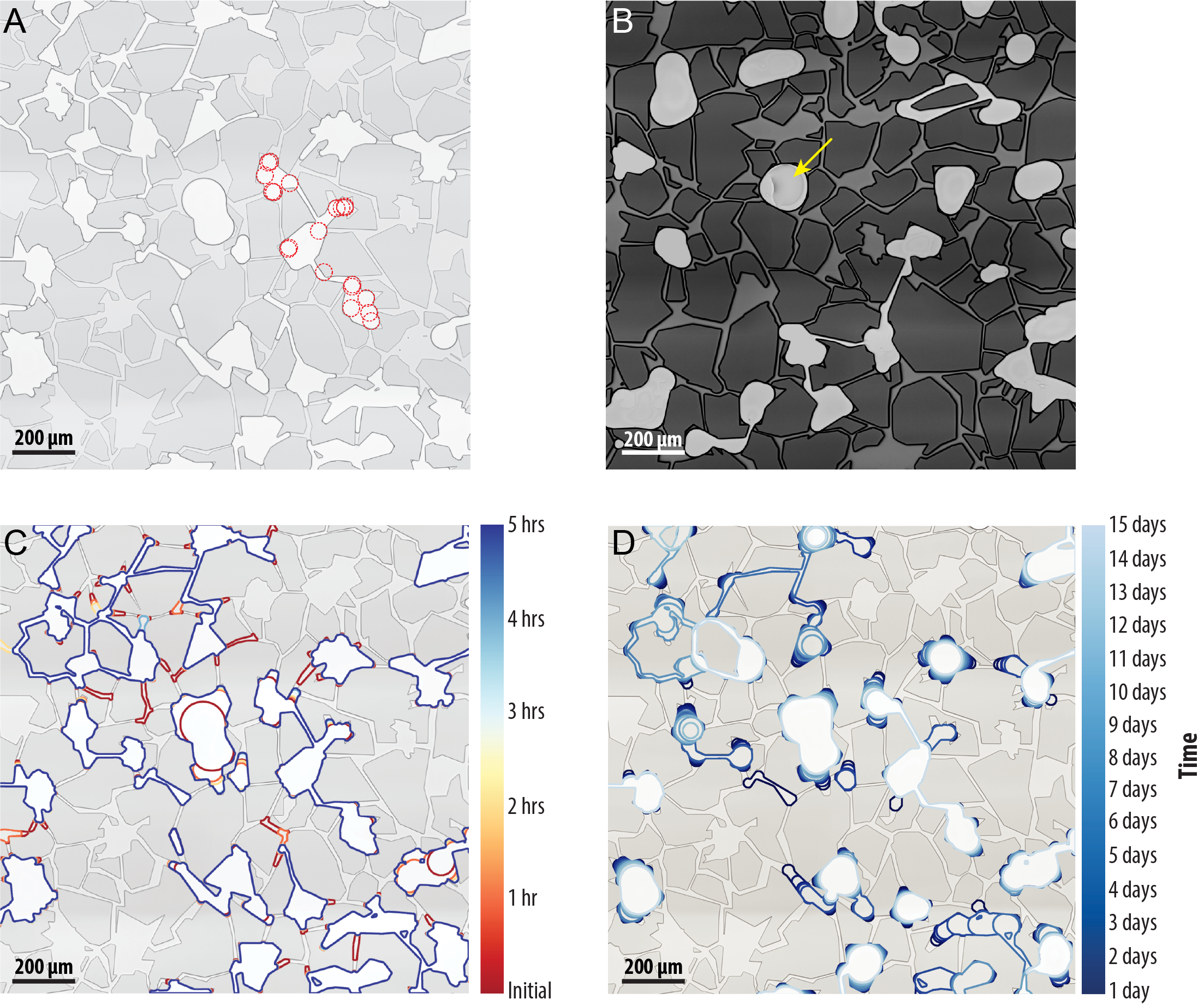}
    \caption{Pore-scale observations of Ostwald ripening in an experiment conducted at 80$^\circ$C. (A) Trapped hydrogen bubbles range in size from single-pore bubbles to ganglia spanning multiple pores. All hydrogen–water interfaces associated with a given bubble exhibit uniform in-plane curvature, as illustrated by least-squares circle fitting (dashed red lines). (B) Small water droplets are visible within some trapped bubbles at the onset of the experiment due to condensation. (C) Evolution of hydrogen–water interfaces for bubbles located near the center of the micromodel, with interface contours color-coded by time. Neighboring bubbles rapidly reach local equilibrium: some grow, others shrink, and some remain unchanged. (D) Long-term evolution of hydrogen–water interfaces for the same bubble population. All bubbles progressively shrink due to diffusive mass transfer toward the left and right boundaries. \label{fig:pore}}
\end{figure*}

\section*{Results}

\noindent\textbf{Pore-scale Observations.} The sizes of trapped hydrogen bubbles span more than two orders of magnitude, ranging from single-pore bubbles to ganglia that extend across multiple pores. Due to spatial confinement, large ganglia possess multiple hydrogen–water interfaces. The ultra-high resolution of our experiments enables accurate measurement of in-plane interface curvature via least-squares circle fitting. We find that all interfaces associated with a given bubble exhibit the same in-plane curvature, although neighboring bubbles display distinct values (Fig.~\ref{fig:pore}A). This uniformity within individual bubbles indicates local capillary equilibrium, consistent with 3D tomography observations of residually trapped CO$_2$ bubbles~\cite{andrew2014pore, garing2017pore}. A survey of the entire bubble population in the micromodel shows that initial in-plane curvatures vary by an order of magnitude, with no apparent correlation to bubble size~(\emph{SI Appendix}, Fig.~S2).

We observe water droplets within some trapped bubbles at the onset of the experiment (Fig.~\ref{fig:pore}B). These droplets arise from condensation triggered by a sudden increase in system pressure during water reinjection, which is carried out to displace hydrogen immediately prior to the experiment. The abrupt pressure rise increases the total pressure inside the hydrogen bubble, but the partial pressure of water vapor cannot adjust rapidly enough via diffusion. As a result, the vapor becomes supersaturated and condenses, preferentially nucleating on surface imperfections within the micromodel. Once formed, these condensed water droplets remain largely unchanged throughout the experiment (\emph{SI Appendix}, Fig.~S3) and have a negligible impact on the ripening process, as demonstrated later.

We examine a 1.5~mm$\times$1.5~mm region near the center of the micromodel in the experiment conducted at 80$^\circ$C with an initial gas saturation of $S_g^\text{init} = 0.49$. Within this domain, some bubbles grow, others shrink, and some remain unchanged, reaching apparent equilibrium within approximately 5 hours (Fig.~\ref{fig:pore}C). This rapid re-equilibration is facilitated by pore-scale heterogeneity, which offers a wide range of metastable configurations. As a result, individual ganglia require only minimal volume adjustments to achieve capillary equilibrium with their neighbors. At this stage, boundary effects have not yet reached the central region. Over a longer timescale of 1 to 15 days, however, we observe that all bubbles gradually shrink due to diffusive mass loss driven by Ostwald ripening toward the left and right boundaries (Fig.~\ref{fig:pore}D). A high-resolution time-lapse video of the experiment is provided in \emph{SI Appendix}, Movie S1.\\

\begin{figure} [t!]
	\centering
	\includegraphics[width=8.7cm]{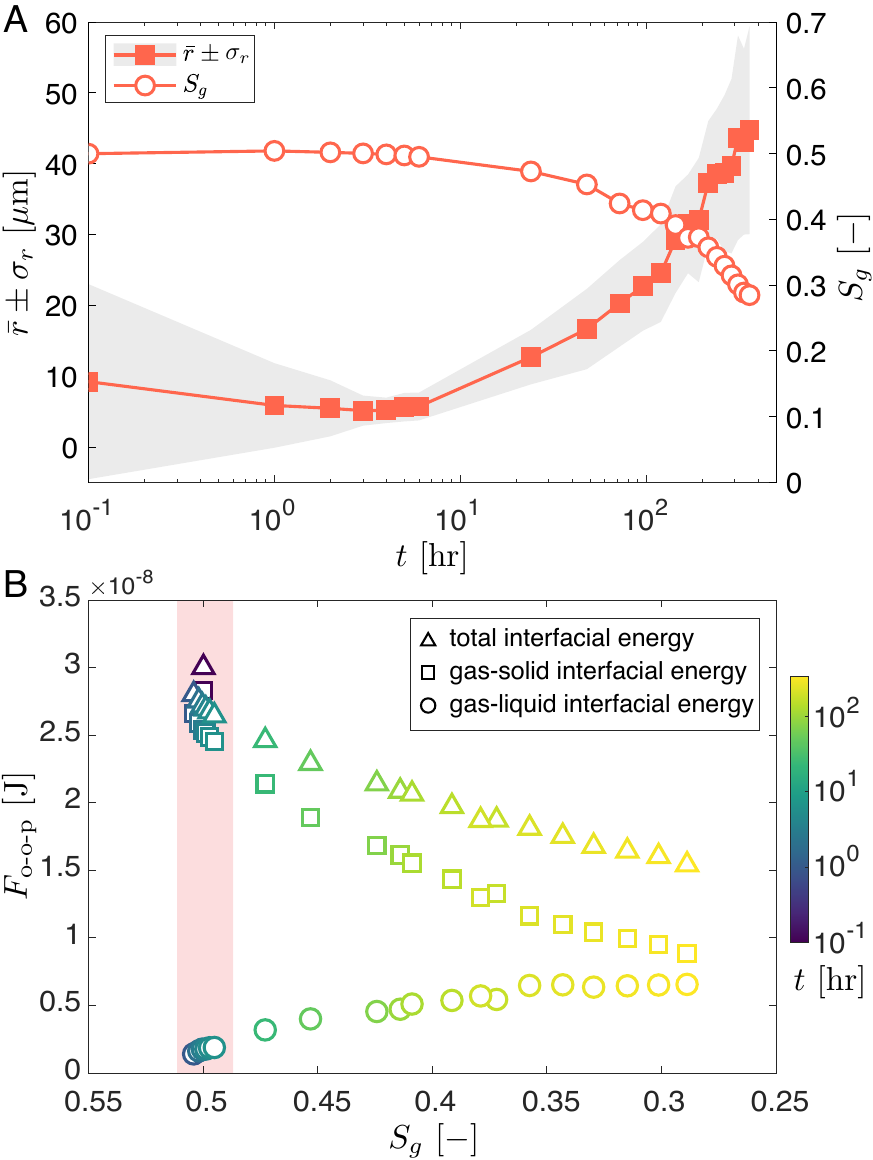}
    \caption{(A) Temporal evolution of the residual gas saturation $S_g$ (circles), average in-plane radius of curvature $\bar{r}$ (squares), and standard deviation of radius of curvature $\sigma_r$ (gray shaded area) for a population of bubbles located in the center of the micromodel in an experiment conducted at 80$^\circ$C. During the first 5 hours, $\sigma_r$ rapidly decreases to near zero, indicating local capillary equilibrium, while $S_g$ and $\bar{r}$ remain relatively constant (i.e. \emph{Stage 1}). Over longer timescales, $S_g$ gradually decreases as bubbles shrink due to diffusive mass transfer toward the boundaries, accompanied by an increase in both $\bar{r}$ and $\sigma_r$ (i.e. \emph{Stage 2}). (B) Evolution of the out-of-plane component of interfacial energy $F_\textrm{o-o-p}$ (triangles) as a function of $S_g$ for the same bubble population. $F_\textrm{o-o-p}$ is decomposed into gas-solid (squares) and gas-liquid (circles) contributions, with symbols color-coded by time. $F_\textrm{o-o-p}$ declines steeply in Stage1 (red shaded area), driven by a precipitous drop in gas–solid interfacial energy, while the gas–liquid interfacial energy increases slightly. These trends persist into Stage~2, albeit at a much slower rate.
 \label{fig:curvature}}
\end{figure}

\noindent\textbf{Two-stage Ripening.} Pore-scale observations of individual bubbles reveal a distinct two-stage evolution during the ripening process. To quantify this behavior, we track the residual gas saturation ($S_g$), the average in-plane radius of curvature ($\bar{r}$), and the standard deviation of radius of curvature ($\sigma_r$) for a population of bubbles within a 1.5~mm-long window that spans the full width of the micromodel. This analysis window is centered within the device, equidistant from the left and right boundaries. Figure~\ref{fig:curvature}A presents these measurements for an experiment conducted at 80$^\circ$C. We find that $\sigma_r$ decreases rapidly at the start of the experiment, reaching a minimum value near zero within five hours, while $S_g$ and $\bar{r}$ remain relatively constant over the same period. The near-zero value of $\sigma_r$ indicates that bubbles in the central region have reached capillary equilibrium, each attaining the same interface curvature. We refer to this rapid equilibration phase as \emph{local ripening (Stage 1)}.

\begin{figure} [h!]
	\centering
	\includegraphics[width=8.7cm]{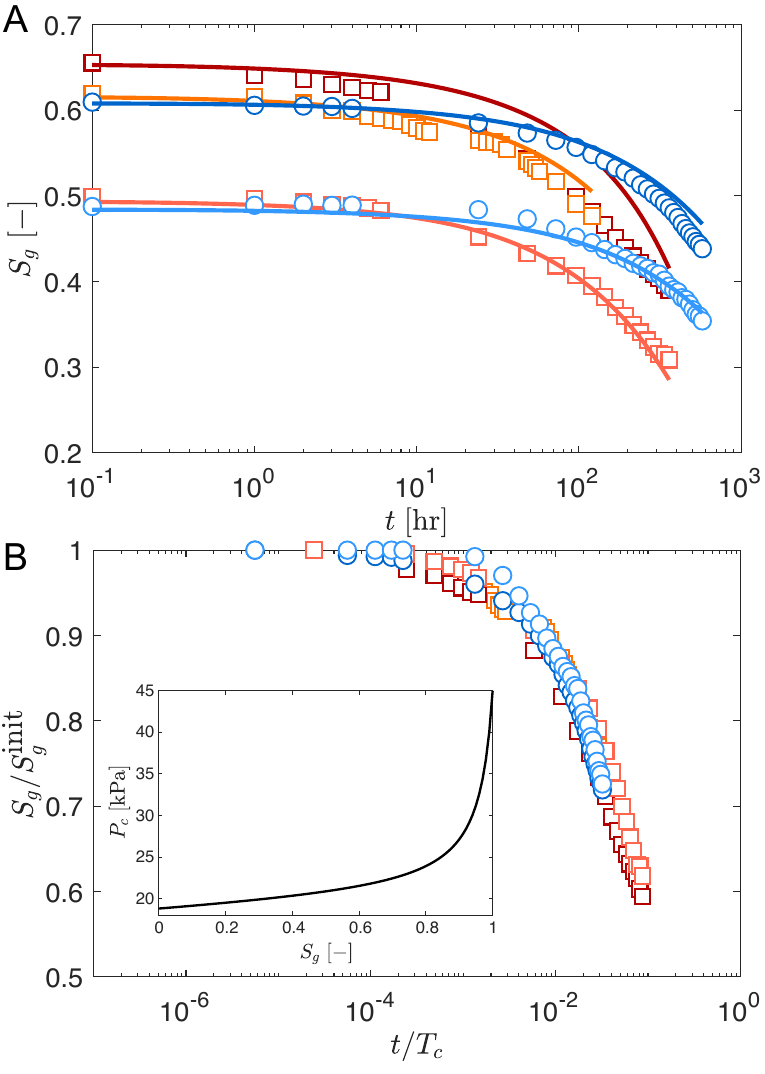}
    \caption{(A) Temporal evolution of the residual gas saturation $S_g$ over the entire micromodel. Square symbols denote experiments conducted at 80$^\circ$C, while circle symbols represent experiments conducted at 40$^\circ$C. Solid lines indicate model predictions (Eqs.~\ref{eq:model_dimensional_chi}). (B) Non-dimensionalizing time by the characteristic diffusion timescale $T_c$ and normalizing $S_g$ by its initial value $S_g^{\text{init}}$ collapses the saturation evolution from different experiments. Inset: $P_c$–$S_g$ relationship computed from pore-scale geometry using the modified pore-morphology method. \label{fig:sat}}
\end{figure}

As the ripening process progresses, bubbles in the central region begin to experience the influence of the boundaries, which exhibit lower chemical potentials due to the larger aperture sizes. This chemical potential gradient drives hydrogen diffusion from the center toward the boundaries, leading to a gradual decrease in $S_g$ as bubbles shrink. This depletion phase is accompanied by a monotonic increase in $\bar{r}$. Meanwhile, $\sigma_r$ rises to nonzero values, as diffusive mass loss disrupts the capillary equilibrium established in \emph{Stage 1}. Shrinking bubbles struggle to maintain metastable configurations with their neighbors, resulting in a broader distribution of curvatures. We refer to this slow, boundary-driven depletion as \emph{global ripening (Stage 2)} (Fig.~\ref{fig:curvature}A). Plotting the evolution of $S_g$ across the entire micromodel reveals the same two-stage ripening process across all experimental conditions, though the distinction between \emph{Stage 1} and \emph{Stage 2} becomes less pronounced, as bubbles near the boundaries begin to lose mass as soon as the experiment starts (Fig.~\ref{fig:sat}A).

To elucidate the thermodynamic driving force underlying the ripening process, we examine the evolution of interfacial energy as a function of gas saturation for the same population of bubbles located at the center of the micromodel. We define the reference energy as that of a fully water-saturated micromodel and assign it a value of zero. Owing to the quasi-two-dimensional geometry of the micromodel, the in-plane component of interfacial energy scales linearly with gas saturation and thus offers limited insight into ripening dynamics. We therefore focus our analysis on the out-of-plane component of interfacial energy given by $F_\text{o-o-p}=(\gamma_\textrm{gs}-\gamma_\textrm{ls})bl_\textrm{gs} + \gamma b \pi l_\textrm{gl}$, where $\gamma_\textrm{gs}$, $\gamma_\textrm{ls}$, and $\gamma$ are the interfacial tensions of the gas-solid, liquid-solid, and gas-liquid interfaces, respectively; $l_\textrm{gs}$ and $l_\textrm{gl}$ are the corresponding interfacial lengths; and $b$ is the micromodel gap thickness. Assuming that water perfectly wets the solid, Young's equation yields $\gamma_\textrm{gs}-\gamma_\textrm{ls}=\gamma$, simplifying the expression to $F_\text{o-o-p} = \gamma b l_\textrm{gs} + \gamma b \pi l_\textrm{gl}$. This formulation represents the total out-of-plane interfacial energy as the sum of the gas–solid and gas–liquid contributions and serves as a thermodynamic indicator of the system's progression toward equilibrium during the ripening process.

As expected, $F_\text{o-o-p}$ decreases monotonically in time as the system evolves toward a lower-energy configuration (Fig.~\ref{fig:curvature}B). The steepest decline occurs during Stage 1, as large bubbles retract from narrow pore throats (Fig.~\ref{fig:pore}C), leading to a rapid decline in gas–solid interfacial energy. This decline, however, is partially offset by a concurrent increase in gas–liquid interfacial energy. These trends persist into Stage 2, albeit at a much slower rate, as diffusive mass loss to the boundaries drives bubbles to further reduce their contact with the solid surface while transitioning toward lower-curvature interfaces (Fig.~\ref{fig:pore}D). The thermodynamic description of multiphase flow in porous media, which couples capillary pressure, saturation, and interfacial area, has been extensively explored~\cite{hassanizadeh1993thermodynamic,mcclure2016influence,ebadi2024}. However, its application to Ostwald ripening remains very limited~\cite{wang2021capillary} and represents a promising direction for future investigation.\\

\noindent\textbf{Continuum Model of Ripening.} To capture the macroscopic dynamics of Ostwald ripening in porous media with pore-scale heterogeneity, we develop a continuum model based on the pore-network formulation of Bueno et al.~\cite{bueno2023ostwald}. The model describes the spatiotemporal evolution of the molar fraction of dissolved gas in the aqueous phase, denoted $\chi$, and is given by
\begin{equation}\label{eq:model_dimensional_chi}
\left\{ \left[ \frac{H}{RT\bar\rho_w} -1 \right] \left[ H \left( \frac{\textrm{d}S_g}{\textrm{d}P_c} \right) \chi + S_g \right] + 1 \right\} \frac{\partial \chi}{\partial t} = \frac{D_w}{\tau} \nabla^2 \chi,
\end{equation}
where $H$ is Henry’s constant in volatility form (with units of pressure per molar fraction), $R$ is the universal gas constant, $T$ is the absolute temperature, $\bar\rho_w$ is the molar density of the aqueous phase, $D_w$ is the molecular diffusion coefficient of the gas in the aqueous phase, and $\tau$ is the tortuosity of the porous medium. Assuming both $\chi$ and $S_g$ vary only along the longitudinal direction, we solve Equation~\ref{eq:model_dimensional_chi} in one dimension (1D). Following the Bruggeman relation~\cite{bruggeman1935berechnung}, we further assume $\tau = \phi^{-1}$, where $\phi$ is the porosity. The term $\textrm{d}S_g/\textrm{d}P_c$ captures the pore-scale coupling between interfacial curvature and gas redistribution, linking saturation changes to capillary pressure gradients. Rather than imposing an empirical capillary pressure–saturation relationship, we compute $\textrm{d}S_g/\textrm{d}P_c$ directly using a modified pore-morphology method (PMM) adapted for ripening, with the micromodel pore geometry as direct input~(\emph{Materials and Methods}). This approach allows us to explicitly encode the influence of pore-scale geometry on macroscopic ripening dynamics~\cite{mehmani2024continuum}. 

We impose Dirichlet boundary conditions on the molar concentration at the left and right inlet channels of the domain, specified by Henry's law as $\chi=(P_w+P_c-P_v)/H$, where $P_w$ is the water pressure (equal to atmospheric pressure in our experiments) and $P_v$ is the saturated vapor pressure of water. Assuming the liquid phase is perfectly wetting to the micromodel, we calculate the capillary pressure $P_c$ at the boundaries using the Young-Laplace equation, $P_c=2\gamma({d^{-1}+b^{-1}})$, where $\gamma$ is the gas-water interfacial tension, $d$ is the constant width of the boundary channels and $b$ is the gap thickness of the micromodel. To relate changes in gas saturation to variations in dissolved concentration, we solve for the capillary pressure as $P_c=H\left(\chi - \chi_0\right)+P_v$, where $\chi_0$ is the equilibrium molar concentration of dissolved gas under ambient temperature and pressure, assuming zero gas–liquid interface curvature. We then invert the capillary pressure–saturation relationship to obtain $S_g$. Remarkably, the model described by Equation~\ref{eq:model_dimensional_chi} captures the experimentally observed evolution of gas saturation without the need for any fitting parameters (Fig.~\ref{fig:sat}A). A full derivation of the model and underlying assumptions is provided in the \emph{SI Appendix}.

Noting that the dimensionless prefactor $H/(RT\bar\rho_w) \gg 1$, due to the low solubility of gases in water, and that
$H \left(\frac{\textrm{d} S_g}{\textrm{d} P_c}\right) \chi  = \left(\frac{\textrm{d} S_g}{\textrm{d} P_c}\right) P_g \gg S_g$,
Equation~\ref{eq:model_dimensional_chi} in 1D simplifies to a diffusion equation:
\begin{equation}\label{eq:model_dimensional_chi_simp}
\frac{\partial \chi}{\partial t}
  ={D_w}{\phi} \left[ H \frac{\bar\rho_g}{\bar\rho_w} \frac{\textrm{d}S_g}{\textrm{d}P_c} + 1 \right]^{-1}\frac{\partial^2 \chi}{\partial x^2},
\end{equation}
where $\bar\rho_g$ is the partial molar density of the gas phase. We define the characteristic timescale as the diffusion timescale, $T_c = L_c^2 / D_{\text{eff}}$, where the characteristic length $L_c$ is set by the length of the micromodel, and the effective diffusivity is given by $D_{\text{eff}} = {D_w}{\phi} \left[ H \frac{\bar\rho_g}{\bar\rho_w} \frac{\textrm{d}S_g}{\textrm{d}P_c} + 1 \right]^{-1}$. Rescaling time $t$ by $T_c$, and residual gas saturation $S_g$ by its initial value, perfectly collapses the saturation evolution curves across different experimental conditions~(Fig.~\ref{fig:sat}B).\\

\noindent\textbf{Large-scale implications.} Saline aquifers are widely regarded as a promising geological setting for large-scale underground gas storage due to their widespread availability and substantial storage capacity~\cite{orr-science-2009, szulczewski-pnas-2012, heinemann2021}. Among these, macroscopically homogeneous saline aquifers---particularly those composed of consolidated sandstones---are especially well-suited owing to their enhanced predictability, containment security, and operational reliability. Notable examples include the Mount Simon Sandstone in the Illinois Basin and the Basal Cambrian Sandstone in Western Canada. Both formations have been extensively characterized and exhibit a high degree of homogeneity over meter to tens-of-meter scales, with relatively uniform porosity and permeability, despite underlying heterogeneity at the pore scale~\cite{medina2011effects, weides2014cambrian}.

Residual trapping in such formations has been investigated through core flooding experiments conducted under reservoir-relevant conditions. These experiments typically involve gas injection into a brine-saturated core (drainage), followed by brine reinjection (imbibition), thereby mimicking the primary processes responsible for residual trapping. When coupled with X-ray computed tomography (CT) imaging, these tests enable direct visualization of the spatial distribution of trapped gas along the core~\cite{zhang2023pore, goodarzi2024trapping}. Analysis of four saturation profiles reveals that the residual gas saturation varies on the centimeter scale and can be reasonably approximated by a sinusoidal function: $S_g^\textrm{init}=\hat{S}_g\sin(2\pi/\lambda~x)+\bar{S}_g$, where $x$ is the longitudinal distance, $\lambda\sim{0.11}-0.63$~cm is the wavelength, $\hat{S}_g\sim{0.03}-0.1$ is the amplitude of saturation variation, and $\bar{S}_g\sim{0.26}-0.46$ is the mean saturation (\emph{SI Appendix}). Furthermore, we apply our modified PMM code to extract $P_c$-$S_g$ relationships from publicly available 3D X-ray CT scans of homogeneous sandstone cores (\emph{SI Appendix, Fig.~S5}). These curves are more appropriate for predicting ripening, as they reflect equilibrium configurations of disconnected ganglia and avoid the viscous effects or hysteresis commonly associated with experimental measurements~(\emph{Materials and Methods}). Noting that $\textrm{d}S_g/\textrm{d}P_c$ is approximately constant over the relevant saturation range ($S_g\in{0.2}-0.6$), Equation~\ref{eq:model_dimensional_chi_simp} can be expressed as an evolution equation for $S_g$:
\begin{equation}\label{eq:model_dimensional_Sg}
\frac{\partial S_g}{\partial t}=D_{\textrm{eff}}\frac{\partial^2S_g}{\partial{x^2}},
\end{equation}
which admits a well-known analytical solution in the absence of a concentration sink (e.g., a fracture). Specifically, the initial variations in ${S}_g$ decay exponentially with a characteristic equilibriation time $t_{\text{eq}}={\lambda^2}/({4\pi^2 D_{\textrm{eff}}})$. 

\begin{figure} [t!]
	\centering
	\includegraphics[width=8.7cm]{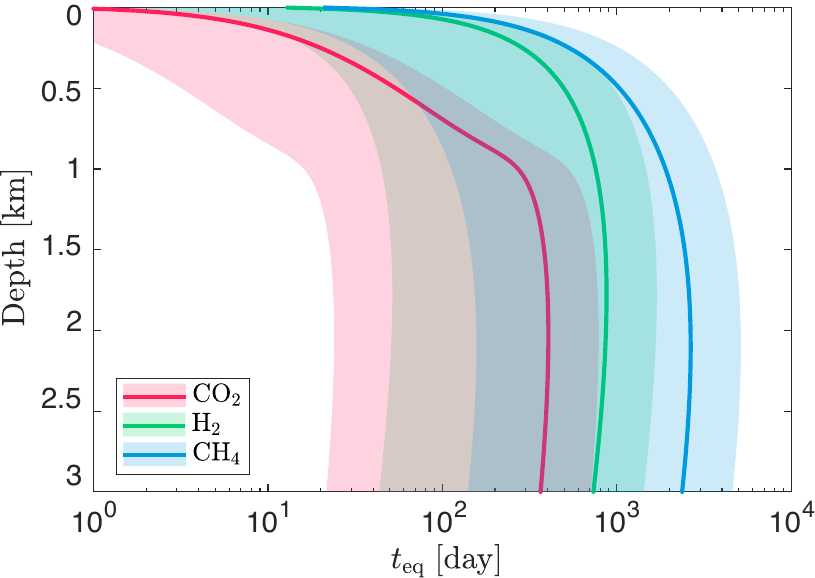}
    \caption{Equilibriation time ($t_{\text{eq}}$) for three gases commonly considered for subsurface storage: CO$_2$ (red),  H$_2$ (green), and CH$_4$ (blue). Shaded regions indicate the range of $t_{\text{eq}}$ values corresponding to the observed variability in the characteristic wavelength of saturation heterogeneity ($\lambda\sim{0.11}-0.63$~cm), as reported in the literature. \label{fig:eq}}
\end{figure}

We estimate $t_{\text{eq}}$ for three gases commonly considered for subsurface storage: H$_2$, CO$_2$, and natural gas (CH$_4$). To account for depth-dependent effects, we develop constitutive models that capture the nonlinear variation of $D_w$, $H$, and $\bar{\rho}_g$ with depth, incorporating geothermal and hydrostatic gradients while neglecting the influence of brine salinity. These constitutive models are validated against published experimental and field data (\emph{SI Appendix}). Across all three gases, $t_{\text{eq}}$ increases with depth, but the change becomes marginal beyond approximately 1 km, where increases in $D_w$ are largely offset by rising $H$, and variations in $\bar{\rho}_g$ are minor. Across all depths, CO$_2$ exhibits the shortest equilibriation time, followed by H$_2$ and then CH$_4$. The equilibriation time also varies substantially with $\lambda$. For a macroscopically homogeneous saline aquifer at a depth of 1 km, with $\phi = 0.2$ and $\lambda \sim 0.11-0.63$ cm as inferred above, the estimated $t_{\text{eq}}$ for CO$_2$ ranges from approximately 20 to 500 days, with a median value of 300 days (Fig.~\ref{fig:eq}). This is substantially shorter than the onset time of convective dissolution in mobile CO$_2$ plumes, typically 10–100 years in similar saline aquifers~\cite{riaz2006onset,pau2010high}. In the absence of sinks such as fractures, ripening-induced equilibration is not expected to compromise residual trapping, as only minimal volume adjustments are required to reach local capillary equilibrium given the broad range of metastable bubble configurations at the pore scale (Fig.~\ref{fig:pore}C). Meanwhile, the estimated $t_{\text{eq}}$ for H$_2$ ranges from approximately 50 to 1,500 days, with a median of 700 days---comparable to the 6–12 month injection–withdrawal cycles characteristic of seasonal underground hydrogen storage, yet orders of magnitude shorter than the million-year timescales required to dissolve an entire column of residually trapped gas via ripening~\cite{blunt2022ostwald}. Depth-dependent ripening arises from the increase in hydrostatic pressure with depth, which elevates the equilibrium concentration of dissolved gas, thereby creating a vertical concentration gradient and sustaining an upward diffusive flux. Our analysis does not account for ripening in the vertical direction.

The results above pertain to local ripening (\emph{Stage 1}). A second implication arises from global ripening (\emph{Stage 2}), also captured by our experimentally validated continuum model (Eq.~\ref{eq:model_dimensional_chi_simp}). Mehmani and Xu~\cite{mehmani2024continuum} solved a linearized form of this model to examine bubble depletion in an infinite, microscopically heterogeneous porous medium, where large bubbles in cavities or fractures act as constant-concentration sinks. Their analysis showed that the mass flux into such sinks may decrease, remain constant, or increase depending on whether ripening proceeds in linear, cylindrical, or radial geometries. The latter two geometries are especially relevant to annular gaps behind wellbore casings and their cement sheath, where ripening can sustain or even accelerate leakage. Because these outcomes are direct mathematical consequences of Equation~\ref{eq:model_dimensional_chi_simp}, now corroborated by experiments, they underscore wellbore integrity as a critical vulnerability in underground gas storage---an issue evidenced by recent explosions and uncontrolled releases linked to wellbore corrosion in underground gas storage facilities~\cite{goodman2022subsurface}.\\

\section*{Conclusions}

Ostwald ripening plays a critical role in governing the long-term fate of residually trapped gas in subsurface gas storage. Through high-resolution microfluidic experiments performed under well-defined boundary conditions, we present the first long-term investigation of Ostwald ripening dynamics within heterogeneous porous media. Our results provide direct experimental evidence that ripening proceeds in two distinct stages: a rapid local equilibration driven by pore-scale heterogeneity, followed by a slower global depletion governed by long-range diffusive mass transfer toward boundary regions with lower chemical potentials (Figs.~\ref{fig:pore}--\ref{fig:curvature}).

Leveraging these insights, we develop and validate a continuum-scale model that couples pore-scale capillary pressure-saturation ($P_c$-$S$) relationships to diffusive mass transport. Crucially, this study is the first to demonstrate that $P_c$-$S$ curves extracted using a simplified pore morphology method adapted for ripening can accurately predict the evolution of gas saturation driven by Ostwald ripening, without the need for fitting parameters (Fig.~\ref{fig:sat}). Applying the validated model to macroscopically homogeneous sandstone aquifers, we estimate characteristic equilibration times ($t_{\text{eq}}$) for CO$_2$, H$_2$, and CH$_4$ under reservoir-relevant conditions. We find that $t_{\text{eq}}$ increases with depth but plateaus beyond approximately 1~km due to compensating thermophysical trends. Among these gases, CO$_2$ equilibrates most rapidly, followed by H$_2$ and CH$_4$ (Fig.~\ref{fig:eq}).

Our findings show that Ostwald ripening occurs significantly faster than the onset of convective dissolution in mobile CO$_2$ plumes during geologic carbon sequestration. In contrast, the characteristic timescale of ripening is comparable to the seasonal injection–withdrawal cycles typical of underground hydrogen storage. Notably, our analysis reveals that the equilibration timescale is sensitive to the spatial wavelength of saturation heterogeneity---a parameter we quantify using existing core-scale experiments. Future studies should focus on how pore geometry, wettability, and flow conditions influence the initial saturation heterogeneity.

Altogether, this study provides a quantitative framework for understanding and predicting Ostwald ripening in subsurface porous media. By directly linking pore-scale geometry to continuum-scale gas transport, our approach establishes a physics-based foundation for assessing the long-term security and efficiency of geological gas storage systems.

\section{Materials and Methods}

\subsection*{Imaging}

We visualize the ripening process using an upright brightfield microscope (BX51, Olympus) equipped with a 10$\times$ objective lens and a scientific CMOS camera (BUC3D-1000C, BestScope) featuring a 10-megapixel sensor. To capture the entire micromodel at ultra-high resolution, we employ a motorized XY stage (EK 14 mot, Marzhauser Wetzlar), which is programmed to traverse the field of view in a serpentine pattern with 35\% overlap between adjacent frames. Each scan acquires 15 columns and 8 rows of images, yielding a total of 120 individual frames, with each frame spanning 1.14 mm~$\times$~0.85 mm. These images are stitched using the Grid/Collection Stitching plugin in Fiji~\cite{preibisch2009globally}, producing a composite image with a spatial resolution of 0.3~$\mu$m/pixel that enables detailed observation of pore-scale ripening dynamics across the entire micromodel. 

To characterize the trapped hydrogen bubbles, we developed a custom image analysis pipeline in MATLAB. First, a dry image of the micromodel is processed using intensity thresholding and morphological operations to extract the pore structure. This step also generates a binary mask that identifies the solid matrix, which is subsequently applied to all time-lapse frames to exclude static grain regions. Gas bubbles are then identified via adaptive thresholding and mask subtraction. Each identified bubble is uniquely labeled and analyzed to determine its centroid position, projected area, pore occupancy number, gas–water and gas–solid interfacial lengths, and Euler characteristic. This enables robust tracking and morphological analysis of individual bubbles throughout the ripening process.

\subsection*{Pore morphology model}

We parameterize the continuum model using a capillary pressure–saturation ($P_c$–$S_g$) relationship derived directly from the binary image of the micromodel pore geometry. Our approach follows the method proposed by Mehmani and Xu~\cite{mehmani2024continuum}, a variant of the classical pore-morphology method (PMM)~\cite{hazlett1995pmm} adapted to capture Ostwald ripening dynamics. Let $I$ denote the binary image, where solid and void regions are represented by 0 and 1, respectively. Our PMM involves two successive morphological operations—erosion followed by dilation—using a circular structuring element $B_r$ of radius $r$. The erosion operation, $I \ominus B_r$, identifies all pixels for which the entire structuring element lies within the void space. The dilation operation, $I \oplus B_r$, subsequently expands these pixels outward by a distance $r$. In this framework, $r$ corresponds to the in-plane radius of curvature of gas–liquid interfaces, and thus to a specific capillary pressure via the Young–Laplace equation. To compute the saturation $S$ associated with $P_c = 2\gamma/r$, we evaluate $(I \ominus B_r) \oplus B_r$ and calculate the ratio of nonzero pixels in the resulting image to the total number of void pixels in $I$. The discrete $P_c$–$S$ data obtained from this modified PMM are then fit to a smooth, differentiable curve~\cite{mehmani2024continuum}, enabling computation of $\mathrm{d}S_g/\mathrm{d}P_c$ in the continuum model~(Eq.~\ref{eq:model_dimensional_chi_simp}).

The key distinction between the classical PMM~\cite{hazlett1995pmm} and the variant of Mehmani and Xu~\cite{mehmani2024continuum} lies in their treatment of disconnected ganglia. Classical PMM excludes isolated clusters to ensure that the nonwetting phase remains connected to an inlet, making it well suited for modeling invasion percolation. In contrast, the ripening-specific variant used here retains disconnected ganglia, consistent with the absence of advective flow and the dominant role of local curvature in establishing capillary equilibrium. As a result, the $P_c$–$S$ relationship for ripening is effectively isotropic and independent of flow direction. In invasion-driven processes, by contrast, the $P_c$–$S$ curve is highly sensitive to inlet geometry: an inlet comprising a single narrow pore can produce a nearly flat $P_c$–$S$ curve, while broader or more connected inlets yield steeper relationships. In some cases, the initial distribution of trapped gas among pores of varying sizes can have a non-trivial influence on the resulting $P_c$–$S$ curve~\cite{mehmani2022pore}. In such cases, PMM must be applied selectively to the initially occupied pores. For the pore geometry considered in this study, however, this sensitivity was found to be minor and is therefore neglected.

\begin{acknowledgments}
\noindent\textbf{Acknowledgements}
\newline
This research is jointly supported by the Natural Sciences and Engineering Research Council of Canada (NSERC) Alliance Grants under ALLRP 592525-23 and ALLRP 567631-24 (for MS, TL, BZ) and the National Science Foundation, United States under Grant No. CBET-2348723 (for NB, ZL, YM).
\end{acknowledgments}

\bibliography{refs}

\end{document}